\documentclass[twocolumn,times,nolinenumbers,tighten]{aastex631}

\usepackage{multirow}
\usepackage{stix}
\usepackage[perpage,para]{footmisc}
\renewcommand{\thefootnote}{${}^\S$}

\shorttitle{Limits on Neutrino Emission from GRB\,221009A}
\shortauthors{IceCube Collaboration}

\begin{document}

\title{Limits on Neutrino Emission from GRB\,221009A from MeV to PeV using the IceCube Neutrino Observatory}

\affiliation{III. Physikalisches Institut, RWTH Aachen University, D-52056 Aachen, Germany}
\affiliation{Department of Physics, University of Adelaide, Adelaide, 5005, Australia}
\affiliation{Dept. of Physics and Astronomy, University of Alaska Anchorage, 3211 Providence Dr., Anchorage, AK 99508, USA}
\affiliation{Dept. of Physics, University of Texas at Arlington, 502 Yates St., Science Hall Rm 108, Box 19059, Arlington, TX 76019, USA}
\affiliation{CTSPS, Clark-Atlanta University, Atlanta, GA 30314, USA}
\affiliation{School of Physics and Center for Relativistic Astrophysics, Georgia Institute of Technology, Atlanta, GA 30332, USA}
\affiliation{Dept. of Physics, Southern University, Baton Rouge, LA 70813, USA}
\affiliation{Dept. of Physics, University of California, Berkeley, CA 94720, USA}
\affiliation{Lawrence Berkeley National Laboratory, Berkeley, CA 94720, USA}
\affiliation{Institut f{\"u}r Physik, Humboldt-Universit{\"a}t zu Berlin, D-12489 Berlin, Germany}
\affiliation{Fakult{\"a}t f{\"u}r Physik {\&} Astronomie, Ruhr-Universit{\"a}t Bochum, D-44780 Bochum, Germany}
\affiliation{Universit{\'e} Libre de Bruxelles, Science Faculty CP230, B-1050 Brussels, Belgium}
\affiliation{Vrije Universiteit Brussel (VUB), Dienst ELEM, B-1050 Brussels, Belgium}
\affiliation{Department of Physics and Laboratory for Particle Physics and Cosmology, Harvard University, Cambridge, MA 02138, USA}
\affiliation{Dept. of Physics, Massachusetts Institute of Technology, Cambridge, MA 02139, USA}
\affiliation{Dept. of Physics and The International Center for Hadron Astrophysics, Chiba University, Chiba 263-8522, Japan}
\affiliation{Department of Physics, Loyola University Chicago, Chicago, IL 60660, USA}
\affiliation{Dept. of Physics and Astronomy, University of Canterbury, Private Bag 4800, Christchurch, New Zealand}
\affiliation{Dept. of Physics, University of Maryland, College Park, MD 20742, USA}
\affiliation{Dept. of Astronomy, Ohio State University, Columbus, OH 43210, USA}
\affiliation{Dept. of Physics and Center for Cosmology and Astro-Particle Physics, Ohio State University, Columbus, OH 43210, USA}
\affiliation{Niels Bohr Institute, University of Copenhagen, DK-2100 Copenhagen, Denmark}
\affiliation{Dept. of Physics, TU Dortmund University, D-44221 Dortmund, Germany}
\affiliation{Dept. of Physics and Astronomy, Michigan State University, East Lansing, MI 48824, USA}
\affiliation{Dept. of Physics, University of Alberta, Edmonton, Alberta, Canada T6G 2E1}
\affiliation{Erlangen Centre for Astroparticle Physics, Friedrich-Alexander-Universit{\"a}t Erlangen-N{\"u}rnberg, D-91058 Erlangen, Germany}
\affiliation{Physik-department, Technische Universit{\"a}t M{\"u}nchen, D-85748 Garching, Germany}
\affiliation{D{\'e}partement de physique nucl{\'e}aire et corpusculaire, Universit{\'e} de Gen{\`e}ve, CH-1211 Gen{\`e}ve, Switzerland}
\affiliation{Dept. of Physics and Astronomy, University of Gent, B-9000 Gent, Belgium}
\affiliation{Dept. of Physics and Astronomy, University of California, Irvine, CA 92697, USA}
\affiliation{Karlsruhe Institute of Technology, Institute for Astroparticle Physics, D-76021 Karlsruhe, Germany }
\affiliation{Karlsruhe Institute of Technology, Institute of Experimental Particle Physics, D-76021 Karlsruhe, Germany }
\affiliation{Dept. of Physics, Engineering Physics, and Astronomy, Queen's University, Kingston, ON K7L 3N6, Canada}
\affiliation{Department of Physics {\&} Astronomy, University of Nevada, Las Vegas, NV, 89154, USA}
\affiliation{Nevada Center for Astrophysics, University of Nevada, Las Vegas, NV 89154, USA}
\affiliation{Dept. of Physics and Astronomy, University of Kansas, Lawrence, KS 66045, USA}
\affiliation{Department of Physics and Astronomy, UCLA, Los Angeles, CA 90095, USA}
\affiliation{Centre for Cosmology, Particle Physics and Phenomenology - CP3, Universit{\'e} catholique de Louvain, Louvain-la-Neuve, Belgium}
\affiliation{Department of Physics, Mercer University, Macon, GA 31207-0001, USA}
\affiliation{Dept. of Astronomy, University of Wisconsin{\textendash}Madison, Madison, WI 53706, USA}
\affiliation{Dept. of Physics and Wisconsin IceCube Particle Astrophysics Center, University of Wisconsin{\textendash}Madison, Madison, WI 53706, USA}
\affiliation{Institute of Physics, University of Mainz, Staudinger Weg 7, D-55099 Mainz, Germany}
\affiliation{Department of Physics, Marquette University, Milwaukee, WI, 53201, USA}
\affiliation{Institut f{\"u}r Kernphysik, Westf{\"a}lische Wilhelms-Universit{\"a}t M{\"u}nster, D-48149 M{\"u}nster, Germany}
\affiliation{Bartol Research Institute and Dept. of Physics and Astronomy, University of Delaware, Newark, DE 19716, USA}
\affiliation{Dept. of Physics, Yale University, New Haven, CT 06520, USA}
\affiliation{Columbia Astrophysics and Nevis Laboratories, Columbia University, New York, NY 10027, USA}
\affiliation{Dept. of Physics, University of Oxford, Parks Road, Oxford OX1 3PU, UK}
\affiliation{Dipartimento di Fisica e Astronomia Galileo Galilei, Universit{\`a} Degli Studi di Padova, 35122 Padova PD, Italy}
\affiliation{Dept. of Physics, Drexel University, 3141 Chestnut Street, Philadelphia, PA 19104, USA}
\affiliation{Physics Department, South Dakota School of Mines and Technology, Rapid City, SD 57701, USA}
\affiliation{Dept. of Physics, University of Wisconsin, River Falls, WI 54022, USA}
\affiliation{Dept. of Physics and Astronomy, University of Rochester, Rochester, NY 14627, USA}
\affiliation{Department of Physics and Astronomy, University of Utah, Salt Lake City, UT 84112, USA}
\affiliation{Oskar Klein Centre and Dept. of Physics, Stockholm University, SE-10691 Stockholm, Sweden}
\affiliation{Dept. of Physics and Astronomy, Stony Brook University, Stony Brook, NY 11794-3800, USA}
\affiliation{Dept. of Physics, Sungkyunkwan University, Suwon 16419, Korea}
\affiliation{Institute of Physics, Academia Sinica, Taipei, 11529, Taiwan}
\affiliation{Dept. of Physics and Astronomy, University of Alabama, Tuscaloosa, AL 35487, USA}
\affiliation{Dept. of Astronomy and Astrophysics, Pennsylvania State University, University Park, PA 16802, USA}
\affiliation{Dept. of Physics, Pennsylvania State University, University Park, PA 16802, USA}
\affiliation{Dept. of Physics and Astronomy, Uppsala University, Box 516, SE-75120 Uppsala, Sweden}
\affiliation{Dept. of Physics, University of Wuppertal, D-42119 Wuppertal, Germany}
\affiliation{Deutsches Elektronen-Synchrotron DESY, Platanenallee 6, 15738 Zeuthen, Germany }

\author[0000-0001-6141-4205]{R. Abbasi}
\affiliation{Department of Physics, Loyola University Chicago, Chicago, IL 60660, USA}

\author[0000-0001-8952-588X]{M. Ackermann}
\affiliation{Deutsches Elektronen-Synchrotron DESY, Platanenallee 6, 15738 Zeuthen, Germany }

\author{J. Adams}
\affiliation{Dept. of Physics and Astronomy, University of Canterbury, Private Bag 4800, Christchurch, New Zealand}

\author[0000-0002-9714-8866]{S. K. Agarwalla}
\altaffiliation{also at Institute of Physics, Sachivalaya Marg, Sainik School Post, Bhubaneswar 751005, India}
\affiliation{Dept. of Physics and Wisconsin IceCube Particle Astrophysics Center, University of Wisconsin{\textendash}Madison, Madison, WI 53706, USA}

\author{N. Aggarwal}
\affiliation{Dept. of Physics, University of Alberta, Edmonton, Alberta, Canada T6G 2E1}

\author[0000-0003-2252-9514]{J. A. Aguilar}
\affiliation{Universit{\'e} Libre de Bruxelles, Science Faculty CP230, B-1050 Brussels, Belgium}

\author[0000-0003-0709-5631]{M. Ahlers}
\affiliation{Niels Bohr Institute, University of Copenhagen, DK-2100 Copenhagen, Denmark}

\author[0000-0002-9534-9189]{J.M. Alameddine}
\affiliation{Dept. of Physics, TU Dortmund University, D-44221 Dortmund, Germany}

\author{N. M. Amin}
\affiliation{Bartol Research Institute and Dept. of Physics and Astronomy, University of Delaware, Newark, DE 19716, USA}

\author{K. Andeen}
\affiliation{Department of Physics, Marquette University, Milwaukee, WI, 53201, USA}

% ...

\date{July 22, 2024}

\collaboration{387}{(IceCube Collaboration)}

\begin{abstract}
Gamma-ray bursts (GRBs) have long been considered a possible source of high-energy neutrinos. While no correlations have yet been detected between high-energy neutrinos and GRBs, the recent observation of GRB\,221009A --- the brightest GRB observed by Fermi-GBM to date and the first one to be observed above an energy of 10 TeV --- provides a unique opportunity to test for hadronic emission. In this paper, we leverage the wide energy range of the IceCube Neutrino Observatory to search for neutrinos from GRB\,221009A. We find no significant deviation from background expectation across event samples ranging from MeV to PeV energies, placing stringent upper limits on the neutrino emission from this source.
\end{abstract}

\keywords{Neutrino telescopes -- Gamma-ray bursts -- Particle astrophysics}

\section{Introduction}\label{sec0}

On October 9th 2022 at 13:16:59.99 UT, the Gamma-ray Burst Monitor (GBM) on board the Fermi satellite triggered and located the exceptionally bright long-duration Gamma-Ray Burst GRB\,221009A~\citep{GCN32636,GCN32642}. This source was also observed by Swift~\citep{GCN32632,GCN32635}, Fermi-LAT~\citep{GCN32637}, INTEGRAL (SPI-ACS), Konus-Wind and triangulated by the InterPlanetary Network~\citep[IPN;][]{GCN32641}. A highlight among the many multiwavelength and multimessenger follow-up searches\footnote{\vspace*{-0.2cm}see \url{https://gcn.gsfc.nasa.gov/} \& \url{http://www.astronomerstelegram.org/}} of GRB\,221009A has been the first detection of $\gamma$-rays above an energy of $10$\,TeV from a GRB as observed by LHAASO~\citep{GCN32677}.

The GBM light curve consists of an initial $17$\,s long pulse after the GBM trigger (T0), followed by an extraordinarily bright episode starting $221.1$\,s after T0. The duration (T90) that contains the central 90\% of emission from the GRB has a length of $(325.8\pm6.8)$\,s. Given that this GRB saturated the Fermi-GBM instrument, it is possible that this is an overestimation of the T90. The time-integrated energy flux in the range 10--1000\,keV is reported as $(2.912 \pm 0.001)\times10^{-2}\,{\rm erg}\,{\rm cm}^{-2}$ with a $1.024$\,s peak photon flux at the level of $(2385\pm3)\,{\rm cm}^{-2}\,{\rm s}^{-1}$, making this the most fluent and intense GRB detected by Fermi-GBM~\citep{GCN32642}.

The GRB was also localized using measurements from the Swift Observatory at right ascension $\alpha=288.2645^\circ$ and declination $\delta=+19.7735^\circ$. The 90\% containment of this measurement, as provided by Swift, is $0.61$ arcseconds~\citep{GCN32632}. The redshift $z\simeq0.151$ of the GRB has been inferred from afterglow emission observed with X-SHOOTER of the Very Large Telescope $11.55$\,hr after T0~\citep{GCN32648} corresponding to a luminosity distance of about $D_L\simeq740$\,Mpc~\citep{ParticleDataGroup:2022pth}.

Long-duration GRBs are triggered by the death of massive stars. The gravitational and/or magnetic energy that is released in these events is the principal power for the observed bright $\gamma$-ray display and has long been speculated to facilitate cosmic-ray acceleration~\citep{Waxman:1995vg,Vietri:1995hs}. The interaction of these cosmic rays with ambient radiation and/or matter could yield broadband neutrino emission that, depending on the mechanism, can even precede the electromagnetic emission. The IceCube Observatory has already reported upper limits on neutrino emission of GRB\,221009A from a Fast Response Analysis~\citep[FRA;][]{GCN32665} searching for TeV--PeV neutrino emission in the periods [T0\,$-$\,1\,hr\,,\,T0\,+\,2\,hr] and T0\,$\pm$\,1d. An independent search by the KM3NeT Observatory did not find significant neutrino emission in the period [T0\,$-$\,50\,s\,,\,T0\,+\,5000\,s] either~\citep{GCN32741}.

In this Letter, we report the results of four searches for neutrino emission coincident with GRB\,221009A using data collected with the IceCube Observatory. These searches cover complementary energy ranges from a few MeV up to the PeV scale, probing neutrino emission from the burst core to high-energy neutrinos produced in external shocks during the GRB.

\section{Neutrinos from Gamma-Ray Bursts}\label{sec1}

One of the leading models for the GRB prompt emission involves a hot and dense plasma (``fireball'') that is initially opaque to radiation and expands under its  pressure. The kinetic energy of the resulting highly relativistic outflow is dissipated via internal shocks, which are expected to form from variations of the bulk Lorentz factor~\citep{Rees:1994nw,Daigne:1998xc,Kobayashi:1997jk}, or via magnetic reconnection in Poynting-flux-dominated outflows~\citep{Meszaros:1996ww}. In both cases, electrons are accelerated and produce the bright $\gamma$-ray prompt display via synchrotron emission in magnetic fields. 

Protons or heavier nuclei entrained in the outflow can also be accelerated by these mechanisms and produce high-energy TeV--PeV neutrinos on collisions with background photons~\citep{Waxman:1997ti,Becker:2005ej,Murase:2005hy,Hummer:2011ms,Bustamante:2014oka,Biehl:2017zlw,Pitik:2021xhb,Ai:2022kvd,Liu:2022mqe,Rudolph:2022dky}. Prior to the prompt emission, internal collisions of protons and neutrons below the photosphere have been considered as a source of GeV neutrino emission~\citep{Murase:2013hh,Bahcall:2000sa,Bartos:2013hf,Murase:2022vqf}. Furthermore, reverse shocks that form during the afterglow phase have been considered sites for efficient cosmic-ray acceleration and would yield extremely high energy neutrinos in the EeV energy range from interactions with optical--UV photons~\citep{Waxman:1999ai,Murase:2007yt,Thomas:2017dft,Murase:2022vqf,Zhang:2022lff}. Detection of these neutrinos would be a compelling indicator of (ultra-)high-energy cosmic-ray acceleration in GRBs.

GRBs could also be promising sites of production of thermal neutrino distributions through different mechanisms, for example, in the standard core-collapse processes, where neutrinos would arrive before the shock breakout~\citep{Kistler:2012as}. Neutrino-dominated accretion flows~\citep{Liu:2017kga,Qi:2021xgl} could form in the case of a supernova with jets, producing neutrinos in the MeV energy range that would arrive either prior to the GRB~\citep{Wang:2007nta,Wei:2019hpd, Liu:2015prx}, or during the $\gamma$-ray emission~\citep{Liu:2015prx}. Finally, the fireball could also be a potential site of neutrino emission with thermal spectra~\citep{Halzen:1996qw}, where these neutrinos are predicted to arrive shortly before the photons in a millisecond burst. 

Which (if any) of these mechanisms leads to detectable neutrino emission is uncertain, so it is imperative to consider different time windows and energy spectra of GRB neutrino emission in dedicated searches. Previous searches of high-energy neutrino emission carried out by IceCube have so far yielded upper limits that imposed constraints on optimistic neutrino emission models~\citep{IceCube:2009xmx,IceCube:2012qza,IceCube:2014jkq,IceCube:2016ipa,LIGOScientific:2017ync,IceCube:2017amx,IceCube:2022rlk}; see also~\cite{Lucarelli:2022ush}. 

\vspace{0.4cm}
\section{The IceCube Observatory}\label{sec2}

The IceCube Observatory~\citep{IceCube:2016zyt} is a multicomponent facility located at the geographic South Pole. Its main component consists of an in-ice array that utilizes one cubic kilometer of the deep ultraclear glacial ice as its detector medium. The fiducial volume is instrumented with 5,160 Digital Optical Modules~\citep[DOMs;][]{IceCube:2008qbc}, each hosting one downward-facing photomultiplier tube (PMT), that register the Cherenkov light emitted by relativistic charged particles passing through the detector~\citep{IceCube:2010dpc}. The DOMs are distributed on 86 readout and support cables (``strings'') and are deployed between 1.45\,km and 2.45\,km below the surface. Most strings follow a triangular grid with a width of 125\,m, evenly spaced over the volume.

Eight strings are placed in the center of the array and are instrumented with a denser DOM spacing and typical interstring separation of 55~m. They are equipped with PMTs with 35\% higher quantum efficiency. These strings, along with the nearest layer of the surrounding standard strings, form the DeepCore low-energy subarray~\citep{IceCube:2011ucd}. While the original IceCube array has a neutrino energy threshold of about 100\,GeV, the addition of the denser infill lowers the energy threshold to about 1\,GeV. An additional component of IceCube is the surface air shower array, IceTop~\citep{IceCube:2012nn}, whose data were not used in the analyses presented in this Letter.

Neutrino interactions with matter in the vicinity of the in-ice array create energetic charged particles that are visible via Cherenkov emission. Several triggers are active in IceCube that are designed for the selection of these and other interesting physics events. The principal challenges of astrophysical neutrino observations with IceCube are the large background generated by cosmic-ray interactions in the atmosphere and reconstruction of neutrino direction. The better the direction of the neutrino is known, the more background from directions not compatible with the GRB can be reduced. At high energies ($\gtrsim$ 100\,GeV), enough light is deposited, in particular by energetic muons that are individually reconstructable in both direction and energy, to provide half-degree typical angular resolution. At lower energies ($\sim$\,10--100\,GeV), only enough light is collected to reconstruct directions within uncertainties of several to tens of degrees, forcing higher reliance on time correlations with the GRB to reject background. At the lowest energies ($\lesssim$ 10\,GeV), individual neutrino interactions are not reconstructable, forcing a complete reliance on timing information.

The analyses described in the following section use different strategies to search for neutrino emission from astrophysical transients of a wide variety of energy and time scales. Each analysis is optimized for the detection of neutrino emission in a different energy range, where it surpasses the sensitivity of other analyses. In combination, this allows us to probe neutrino emission of GRB\,221009A over nine orders of magnitude in neutrino energy, from MeV to PeV.

\begin{deluxetable}{cccc}
\tablenum{1}\centering
\renewcommand{\arraystretch}{1.1}
\tablecaption{Models for the time-integrated neutrino flux $F(E)$ and energy ranges probed by different datasets. While the analyses based on GFU, GRECO and ELOWEN test broad power-law fluxes the analysis based on SNDAQ probes peaked quasi-thermal fluxes motivated by core-collapse supernovae.\label{tab:energyrange}} 
\tablehead{
\makebox[1.5cm][c]{Dataset} & \multicolumn{3}{c}{Time-Integrated Neutrino Flux Model}\\
& \multicolumn{3}{c}{Power-law : $F_{\nu+\bar{\nu}}(E)\propto E^{-\gamma}$ for $E_{\rm min} \leq E \leq E_{\rm max}$}\\& \makebox[1.2cm][c]{$\gamma$} & \makebox[2cm][c]{$E_{\rm min}$} & \makebox[2cm][c]{$E_{\rm max}$}}
\startdata
\multirow{4}{*}{GFU${}^\star$}& 1.5 & $6.8$\,TeV & $9.9$\,PeV\\
 & 2.0 & $0.83$\,TeV & $0.96$\,PeV\\
 & 2.5 & $0.23$\,TeV & $0.086$\,PeV\\
 & 3.0 & $0.13$\,TeV & $0.013$\,PeV\\
 \hline
\multirow{4}{*}{GRECO${}^\star$}& 1.5 & $40$\,GeV & $1.5$\,TeV\\
 & 2.0 & $26$\,GeV & $1.2$\,TeV\\
 & 2.5 & $15$\,GeV & $0.70$\,TeV\\
 & 3.0 & $11$\,GeV & $0.35$\,TeV\\
 \hline
ELOWEN & 2.0\,$-$\,3.0 & $0.5$\,GeV & $5.0$\,GeV\\[0.05cm]
\hline
 & \multicolumn{3}{c}{Quasi-thermal : $F_{\bar \nu_e}(E) \propto E^2\exp(-3E/\langle E\rangle)$}\\
SNDAQ& \multicolumn{3}{c}{$E\simeq \langle E\rangle \simeq (10-20)\,{\rm MeV}$} \\[0.05cm]
\enddata
\tablecomments{\,${}^\star$Central 90\% energy range depends on spectral index $\gamma$.}
\end{deluxetable}

\section{Analyses and Results}\label{sec3}

In the following, we summarize the methodology and results of four complementary IceCube analyses of neutrino emission of GRB\,221009A. Three of these are based on neutrino event samples that have been developed for previous IceCube analyses: the Gamma-ray Follow-Up (GFU) sample has been optimized for IceCube's real-time program~\citep{IceCube:2016xci}; the GeV Reconstructed Events with Containment for Oscillations (GRECO) sample was originally developed for oscillation studies with atmospheric neutrinos~\citep{IceCube:2019dqi} and has since been used for low-energy neutrino transients~\citep[GRECO Astronomy;][]{IceCube:2020qls}; the Extremely Low-Energy (ELOWEN) event sample was developed for the search of low-energy neutrino emission during solar flares~\citep{IceCube:2021jwt}. In addition, IceCube is able to observe a significant burst of MeV neutrinos through an increased level of single-photon hit rates in DOMs across the entire detector. This search is based on a real-time data stream called the Supernova Data Acquisition (SNDAQ) that was developed to detect MeV neutrinos from supernovae~\citep{IceCube:2011cwc}.

Table~\ref{tab:energyrange} summarizes the test hypotheses for the time-integrated neutrino flux $F(E) \equiv d^2N/dE/dA$ (neutrinos per area $A$ and energy $E$) and corresponding neutrino energy scales accessible by these different datasets. The energy-dependent effective areas from the direction of GRB\,221009A can be found in Appendix~\ref{sec:appendixA}. For the GFU and GRECO datasets the energy interval $[E_{\rm min},E_{\rm max}]$ corresponds to the central 90\% energy range of predicted events for a power-law emission with spectral index $\gamma$ at the location of GRB\,221009A; harder (softer) indices will shift this energy range to higher (lower) values. Table~\ref{tab:energyrange} shows the wide MeV--PeV bandwidth covered in the following analyses. 

\begin{figure*}[t]
\centering
\includegraphics[width=0.7\linewidth]{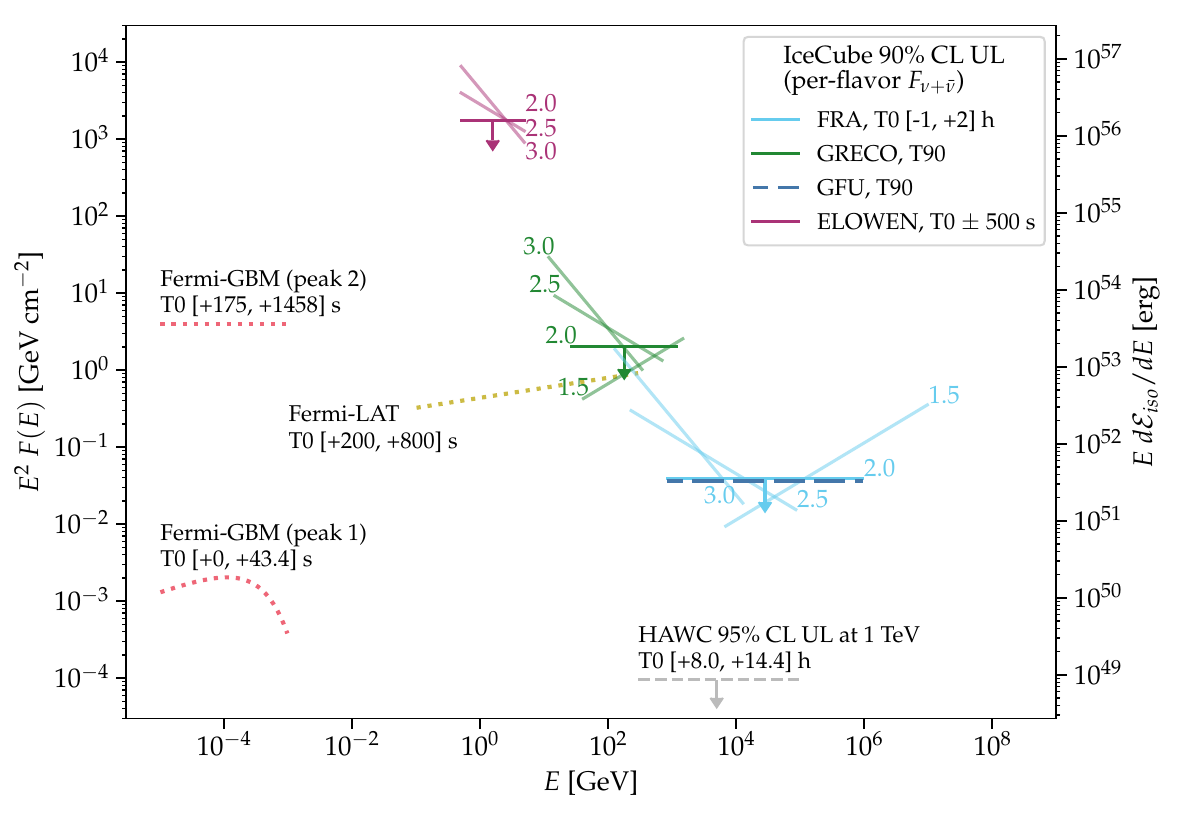}\\[-0.2cm]
\caption{Gamma-ray observations and $F_{\nu+\bar{\nu}}$ upper limits on the time-integrated neutrino flux of GRB\,221009A. We show the $\gamma$-ray observations from Fermi-GBM~\citep{GCN32642} and Fermi-LAT~\citep{GCN32637} as well as upper limits from HAWC~\citep{GCN32683}. The Fermi-GBM result covering the prompt phase (``peak 2'') had no reported spectral fit, so it is shown here at $\gamma=2.0$ for visualization purposes. The upper limits on the time-integrated neutrino flux are shown for various spectral indices as indicated by the numbers. The right axis shows the differential isotropic equivalent energy $d\mathcal{E}_{\rm iso}/dE$.}
\label{fig:fluence_ULs}
\end{figure*}

\subsection{Fast Response Analysis: > 100 GeV}

The FRA~\citep{IceCube:2020mzw, IceCubeCollaboration:2022fxl} is a framework established to rapidly follow up interesting transients or multimessenger events in real time with the IceCube Neutrino Observatory. The FRA requires a real-time data stream, with low-latency access to data from the South Pole, in order to follow up interesting transients in real time. This GFU sample uses events triggered by causally connected single-photon hits on eight or more neighboring IceCube DOMs and employs a faster event reconstruction at the expense of some angular resolution available for offline reconstructions~\citep{IceCube:2016xci}. At final filter level, the stream has an all-sky rate that varies between approximately 6 and 7\,mHz owing to seasonal variations~\citep{IceCube:2020mzw}. Figure~\ref{fig:effective_area} in Appendix~\ref{sec:appendixA} shows the effective area of the GFU sample for the direction of GRB\,221009A.

The FRA uses an unbinned maximum likelihood method to identify significant neutrino emission following an $E^{-2}$ spectrum on top of background events. The method has been used in other searches for short-timescale neutrino emission from potential sources~\citep{IceCube:2019acm, Abbasi:2022rbd, IceCube:2022rlk}. We initiated an FRA of GRB\,221009A in response to the observations reported by Swift and Fermi-GBM using the time windows {\it (a)} [T0\,$-$\,1\,hr\,,\,T0\,+\,2\,hr] (3 hours in total) and {\it (b)} T0\,$\pm$\,1\,d (2 days in total) to cover extended emission periods before and after the Fermi-GBM trigger, as summarized in~\cite{GCN32665}. 

In addition to this initial real-time FRA, we also searched for neutrino emission using the same data sample for an extended set of time windows following the method of a previous archival search for neutrinos coincident with GRBs~\citep{IceCube:2022rlk}. This analysis used three time windows in order to search for neutrinos {\it (c)} during the T90 phase as given by Fermi-GBM, {\it (d)} during the period [T0\,$-$\,200\,s\,,\,T0\,+\,2000\,s] to include times corresponding to the high-energy photons observed by LHAASO~\citep{GCN32677} as well as potential precursor emission, and {\it (e)} during the extended period [T0\,$-$\,1\,d\,,\,T0\,+\,14\,d] covering precursor-to-afterglow emission. We scanned over a circle of radius of $1^\circ$, centered on the location of GRB\,221009A in order to find the location with highest significance, which is consistent with the previous method. Both these analyses assumed an $E^{-2}$ spectrum as in the initial FRA.

None of these analyses --- the initial FRA and its application to an extended set of time windows --- found indications for neutrino emission above background expectations, with $p$-values being close to 1. These results allow us to set upper limits on the energy-scaled time-integrated neutrino flux, $E^2F_{\nu+\bar{\nu}}(E)$, which are summarized in Tab.~\ref{tab:fluence_limits} at a reference energy $E_0=100\,{\rm TeV}$. In the case of the standard FRA time windows of 3 hours and 2 days, we provide upper limits for four different power-law indices $\gamma$ based on Monte Carlo injections following $F_{\nu+\bar{\nu}}(E) \propto E^{-\gamma}$. 

Figure~\ref{fig:fluence_ULs} shows the upper limits on the time-integrated neutrino flux for the T90 phase and the time window [T0\,$-$\,1\,hr\,,\,T0\,+\,2\,hr]. These upper limits can be translated into limits on the differential isotropic equivalent energy $d\mathcal{E}_{\rm iso}/dE$ of GRB\,221009A at redshift $z$ and luminosity distance $D_L$ via the relation $E^2 F_{\nu+\bar{\nu}}(E) = Ed\mathcal{E}_{\rm iso}/dE\times(1+z)/(4 \pi D_L^2)$. This quantity is indicated by the right axis in Fig.~\ref{fig:fluence_ULs}. 

\begin{deluxetable*}{ccc|ccccc}
\tablenum{2}
\renewcommand{\arraystretch}{1.1}
\tablecaption{Upper limits (UL) on the time-integrated neutrino flux of GRB\,221009A for different time windows${}^\sharp$. The three analyses based on GFU, GRECO and ELOWEN assume a per-flavor time-integrated power-law flux $F_{\nu+\bar{\nu}}(E)\propto E^{-\gamma}$ with variable spectral index $\gamma$. The results are given for the energy-scaled time-integrated per-flavor neutrino flux, $E^2F_{\nu+\bar{\nu}}(E)$, at a reference energy $E_0$ depending on the dataset. The analysis based on the SNDAQ assumes an electron antineutrino spectrum following a quasi-thermal spectrum peaked at $\langle E\rangle=15$\,MeV. Results are shown for the total and peak time-integrated $\bar\nu_e$ flux. 
\label{tab:fluence_limits}}
\tablewidth{0pt}
\tablehead{
\makebox[1.5cm][c]{Dataset} & \multicolumn{2}{c|}{\makebox[3.cm][c]{Time Window \& Index${}^\sharp$}} & \multicolumn{5}{c}{90\% C.L.~Upper Limits (ULs) on the Time-integrated Neutrino Flux $F(E)$}\\
& & & \multicolumn{5}{c}{Power-law $F_{\nu+\bar{\nu}}(E)\propto E^{-\gamma}$: per-flavor ULs show $E^2 F_{\nu+\bar{\nu}}(E)$ [GeV cm$^{-2}$] at $E_0$}\\
& & & \makebox[1.5cm][c]{$E_0$} & \makebox[2cm][c]{$\gamma=1.5$} & \makebox[2cm][c]{$\gamma=2.0$} & \makebox[2cm][c]{$\gamma=2.5$} & \makebox[2cm][c]{$\gamma=3.0$}}
\startdata
\multirow{5}{*}{GFU} & {[T0\,$-$\,1\,hr\,,\,T0\,+\,2\,hr]} & {\it (a)} & \multirow{5}{*}{$100$~TeV} & 0.0359  & 0.0393${}^\star$ & 0.0143  & 0.00240 \\
& {T0\,$\pm$\,1\,d}             &  {\it (b)} & & 0.0370  & 0.0410${}^\star$ & 0.0176  & 0.00345 \\
& T90 phase &  {\it (c)} & & \nodata & 0.0364 & \nodata & \nodata \\
& {[T0\,$-$\,200\,s\,,\,T0\,+\,2000\,s]} &  {\it (d)} & & \nodata & 0.0369               & \nodata & \nodata \\
& {[T0\,$-$\,1\,d\,,\,T0\,+\,14\,d]}   &  {\it (e)} & & \nodata & 0.0471               & \nodata & \nodata \\
\hline
\multirow{2}{*}{GRECO} & T90 phase & {\it (c)}  & \multirow{2}{*}{$1$~TeV}& 2.104 & 2.030 & 1.122 & 0.348  \\
& {[T0\,$-$\,200\,s\,,\,T0\,+\,2000\,s]} & {\it (d)} & & 2.774 & 2.676 & 1.480 & 0.458  \\
\hline
\multirow{2}{*}{ELOWEN} & T0\,$\pm$\,500\,s & {\it (f)} & \multirow{2}{*}{$1$~GeV} & \nodata & $1.8 \times 10^3$ & $2.9 \times 10^3$ & $0.47\times 10^4$ \\
& {[T0\,$-$\,200\,s\,,\,T0\,+\,2000\,s]} & {\it (d)}  & &\nodata & $2.6\times 10^3$ & $0.43 \times 10^4$ & $0.67\times 10^4$\\[0.05cm]
\hline\hline
& & & \multicolumn{5}{c}{Quasi-thermal $F_{\bar \nu_e}(E) \propto E^2\exp(-3E/\langle E\rangle)$ : ${\bar \nu_e}$ UL on total and peak flux}\\
& & & $\langle E\rangle$ & \multicolumn{2}{c}{Total ${\bar \nu_e}$ Flux [cm$^{-2}$]} & \multicolumn{2}{c}{$E^2 F_{\bar \nu_e}(E)$ [GeV cm$^{-2}$] at $\langle E\rangle$}\\[0.1cm]
\hline
\multirow{6}{*}{SNDAQ} & {[T0\,$-$\,100\,s\,,\,T0]} & {\it (g)} & \multirow{6}{*}{$15$~MeV} & \multicolumn{2}{c}{$7.98\times 10^{8}$} & \multicolumn{2}{c}{$8.05\times10^6$} \\ 
& {[T0\,$-$\,1\,s\,,\,T0]}  & {\it (h)} & & \multicolumn{2}{c}{$1.81\times10^{9}$} & \multicolumn{2}{c}{$1.82\times10^7$}\\
& {[T0\,,\,T0\,+\,17\,s]}  & {\it (i)} & & \multicolumn{2}{c}{$8.00\times 10^{8}$} & \multicolumn{2}{c}{$8.07\times 10^{6}$}\\
& {[T0\,+\,18\,s\,,\,T0\,+\,174\,s]} & {\it (j)} & & \multicolumn{2}{c}{$3.08\times 10^{8}$} & \multicolumn{2}{c}{$3.11\times 10^{6}$}\\
& {[T0\,+\,174\,s\,,\,T0\,+\,175\,s]} & {\it (k)} & & \multicolumn{2}{c}{$1.35\times 10^{9}$} & \multicolumn{2}{c}{$1.36\times 10^{7}$}\\
& {[T0\,+\,175\,s\,,\,T0\,+\,547\,s]}& {\it (l)} & & \multicolumn{2}{c}{$4.00\times 10^{8}$} & \multicolumn{2}{c}{$4.03\times 10^{6}$}\\[0.05cm]
\enddata
\tablecomments{\,${}^\sharp$The different time windows are discussed in section~\ref{sec3} and are referenced by their inline text indices.}
\tablecomments{\,${}^\star$Values corresponding to the real-time FRA results published in~\cite{GCN32665}.}
\end{deluxetable*}

\vspace{-1cm}
\subsection{GRECO Astronomy Analysis: 10--1000 GeV}

This analysis focuses on the low-energy GRECO Astronomy dataset, which is optimized for neutrinos between 10\,GeV and 1\,TeV~\citep{IceCube:2022lnv}. This sample is based on events with at least three coincident hits in DeepCore DOMs and has an average event rate of about 4\,mHz at final filter level~\citep{IceCube:2019dqi}. The effective area for GRB\,221009A is shown in Fig.~\ref{fig:effective_area}. In order to test for neutrino emission from GRB\,221009A, we use an extended unbinned maximum-likelihood outlined in~\cite{IceCube:2022rlk} combined with a spatial prior to account for the source localization uncertainty. The signal hypothesis assumes an $E^{-2.5}$ neutrino spectrum with equal contributions from all flavors.

To account for the fact that angular uncertainties for the low-energy events used in the GRECO Astronomy dataset are relatively large, a Kent distribution~\citep{Kent82} is assumed for the spatial likelihood. The localization prior is a $1^\circ$ radius top-hat distribution centered at the localization provided by Swift. This is done for computational reasons, and the size of the prior does not affect the result given that angular uncertainties used in GRECO Astronomy are relatively large. 

The analysis of GRB\,221009A focused on two time windows, matching those of other analyses at different energy ranges. We searched for neutrino emission ${\it (c)}$ coincident with the central 90\% emission from the GRB as reported as the T90 phase by Fermi-GBM and ${\it (d)}$ the 2,200 second window [T0\,$-$\,200\,s\,,\,T0\,+\,2000\,s] including potential precursor emission and coinciding with high-energy photons observed by LHAASO. Though even longer time windows are desirable too, this is prevented by computational constraints.

We did not find significant deviations from a background distribution in either of the two time windows, with $p$-values being close to 1 in this analysis. The corresponding upper limits on the energy-scaled time-integrated neutrino flux, $E^2 F_{\nu+\bar{\nu}}(E)$, at a reference energy of $E_0=1$~TeV for each time window and four different spectral indices can be found in Tab.~\ref{tab:fluence_limits}. Similar to the previous analysis, we provide upper limits for different power-law indices $\gamma$ derived from the corresponding Monte-Carlo injections. Figure~\ref{fig:fluence_ULs} shows the upper limits on the time-integrated neutrino flux for the T90 phase.

\vspace{-0.2cm}
\subsection{ELOWEN Analysis: 0.5--5 GeV}

The selection of events below 10\,GeV at IceCube is challenging owing to the presence of large atmospheric backgrounds and the lack of directional reconstruction in this energy range. However, relying on temporal coincidence makes it possible to search for potential neutrino emission in the GeV energy range. Here we follow the methods used in a previous analysis in this energy range \citep{IceCube:2021ddq}, which are briefly summarized here. Two short time windows were chosen based on archival electromagnetic data from GRBs~\citep{Baret:2011tk} while minimizing the impact of the background for our neutrino search: {\it (f)} a 1,000 second window T0\,$\pm$\,500\,s centered on the Fermi-GBM trigger and {\it (d)} the 2,200 second window [T0\,$-$\,200\,s\,,\,T0\,+\,2000\,s] also used in other analyses. 

Because of the reliance on timing in this analysis, we require a precise model of the background event rate as the key component of the search. To determine this, we examined a large number of time windows of the same lengths during time periods where no transient events were detected (GRBs, gravitational waves, or --- for the longer window --- classical novae). The rates in these windows, as well as in the 8-hour period immediately before the GRB, were statistically compatible with the expected background data rate of 0.2\,Hz, validating our background model.

Like GRECO, the dataset is based on events with at least three coincident hits in DeepCore DOMs. The filtering process for ELOWEN, which is explained in more detail in \cite{IceCube:2021jwt}, keeps 40\% of the initial sample of neutrinos following an $E^{-2}$ spectrum from 0.5\,GeV to 5\,GeV, simulated with GENIE 2.8.6~\citep{Andreopoulos:2009rq}, while reducing the background from atmospheric muons and detector noise~\citep{IceCube:2010dpc,IceCube:2008qbc,MLarson} by $>$99.9\%. We show the ELOWEN effective area in comparison to the previously discussed samples in Fig.~\ref{fig:effective_area}.

The significance of any excess observed during the on-time phase can then be calculated by comparison to the off-time phase with the methods described in \cite{Li:1983fv}. None of the tested time windows showed indications for neutrino emission in the 0.5\,$-$\,5\,GeV energy range; we find $p$-values larger than $0.7$. The corresponding upper limits on the energy-scaled time-integrated neutrino flux, $E^2F_{\nu+\bar{\nu}}(E)$, at a reference energy of $E_0=1$\,GeV can be found in Tab.~\ref{tab:fluence_limits} for three different spectral indices of the simulated power-law spectrum. Figure~\ref{fig:fluence_ULs} shows the upper limits on the time-integrated flux for the 1,000 second time window.

\vspace{-0.1cm}
\subsection{MeV Neutrino Burst Analysis: < 1 GeV}

IceCube can observe a significant burst of sub-GeV neutrinos through an increased level of single-photon hit rates on the PMTs across the entire detector, predominantly from the reaction $\bar\nu_e + p \to e^+ +n$, compared to the uncorrelated single-PMT noise rate at the level of about $0.5$~kHz~\citep{IceCube:2010dpc}. In this analysis, we use the SNDAQ data stream designed for the detection of supernovae~\citep{IceCube:2011cwc} to search for MeV neutrinos from GRB\,221009A. We divide the search into six time windows covering from 100\,s prior to T0 up to the end of T90 for a total of 647 seconds. These time windows, shown at the bottom of Tab.~\ref{tab:fluence_limits}, were selected to search for MeV neutrino emission {\it (g)} 100\,s and {\it (h)} 1\,s prior to the precursor, {\it (i)} during the precursor phase, {\it (j)} during the phase between precursor and prompt emission, {\it (k)} 1\,s prior to the prompt phase, and {\it (l)} during the T90 phase. To find the excess rates in the different time windows, we use a sliding 1\,s search window to identify the highest rate during each of the longer time periods. For the shortest time windows spanning 1\,s, we instead use a 0.5\,s sliding search window. 

To calculate the significance of the observed number of hits at the time of the GRB compared to what is expected from the background, we repeat the search method described above for each time window on 2 weeks of off-time data: 1 week prior to the GRB and 1 week after. The upper limit on the observed hits is obtained using the method of \cite{Feldman:1997qc} for a 90\% C.L. These hit upper limits are then used to obtain an upper limit on the time-integrated flux through A Supernova Test Routine for IceCube Analysis~\citep[ASTERIA;][]{Griswold2020}. ASTERIA is a fast supernova neutrino simulation designed to model the detector response for IceCube.

The neutrino spectra of core-collapse supernovae are expected to evolve in time. For simplicity, all tested time windows in this analysis assume the emission of electron antineutrinos ($\bar\nu_e$) following a generic quasi-thermal spectrum inferred from simulations of core-collapse supernovae~\citep{Keil:2002in} approximated as a Maxwell--Boltzmann distribution $F_{\bar\nu_e}(E) \propto E^2\exp(-3E/\langle E\rangle)$ with mean energy of $\langle E\rangle=15$\,MeV~\citep{Tamborra:2012ac}. Since the on-time data are consistent with the background for all six time windows ($p>0.3$), we set a 90\% C.L.~upper limit for the total time-integrated flux of electron antineutrinos and the energy-scaled time-integrated flux, $E^2F_{\bar\nu_e}(E)$, at the peak $\langle E\rangle$ for each time window in Tab.~\ref{tab:fluence_limits}. 

\vspace{0.2cm}
\section{Conclusions}\label{sec4}

Our searches did not find significant neutrino emission in an energy range from MeV to PeV, before, during, or after the electromagnetic emission, resulting in upper limits on the time-integrated neutrino flux of GRB\,221009A. These upper limits allow constraints on neutrino emission models from this source, as discussed by \cite{Ai:2022kvd}, \cite{Murase:2022vqf}, \cite{Liu:2022mqe}, and \cite{Rudolph:2022dky}. Previous IceCube searches for the joint neutrino emission of GRBs showed that high-energy neutrino emission during the prompt phase (or during $10^4$\,s) of GRBs is limited to $\leq1\%$ (or $\leq24\%$) of the high-energy diffuse neutrino flux observed by IceCube~\citep{IceCube:2022rlk}.
 
Future upgrades and proposed extensions of IceCube could provide a crucial step toward neutrino detection from GRBs. The IceCube-Upgrade~\citep{Ishihara:2019aao} will enhance the detection of 1-10 GeV neutrino events with respect to the GRECO Astronomy and ELOWEN samples and will improve the angular reconstruction above $10$\,GeV. The proposed multicomponent extension IceCube-Gen2~\citep{IceCube-Gen2:2020qha} would significantly increase the sensitivity to high-energy neutrino transients, {\it e.g.},~increasing the observable volume within our universe by one order of magnitude for TeV--PeV neutrinos compared to IceCube.

\vspace{0.2cm}
\section*{Acknowledgments} 

The IceCube Collaboration acknowledges Eric Burns for valuable discussions of public Fermi-GBM results. The authors gratefully acknowledge the support from the following agencies and institutions: USA {\textendash} U.S. National Science Foundation-Office of Polar Programs,
U.S. National Science Foundation-Physics Division,
U.S. National Science Foundation-EPSCoR,
Wisconsin Alumni Research Foundation,
Center for High Throughput Computing (CHTC) at the University of Wisconsin{\textendash}Madison,
Open Science Grid (OSG),
Advanced Cyberinfrastructure Coordination Ecosystem: Services {\&} Support (ACCESS),
Frontera computing project at the Texas Advanced Computing Center,
U.S. Department of Energy-National Energy Research Scientific Computing Center,
Particle astrophysics research computing center at the University of Maryland,
Institute for Cyber-Enabled Research at Michigan State University,
and Astroparticle physics computational facility at Marquette University;
Belgium {\textendash} Funds for Scientific Research (FRS-FNRS and FWO),
FWO Odysseus and Big Science programmes,
and Belgian Federal Science Policy Office (Belspo);
Germany {\textendash} Bundesministerium f{\"u}r Bildung und Forschung (BMBF),
Deutsche Forschungsgemeinschaft (DFG),
Helmholtz Alliance for Astroparticle Physics (HAP),
Initiative and Networking Fund of the Helmholtz Association,
Deutsches Elektronen Synchrotron (DESY),
and High Performance Computing cluster of the RWTH Aachen;
Sweden {\textendash} Swedish Research Council,
Swedish Polar Research Secretariat,
Swedish National Infrastructure for Computing (SNIC),
and Knut and Alice Wallenberg Foundation;
European Union {\textendash} EGI Advanced Computing for research;
Australia {\textendash} Australian Research Council;
Canada {\textendash} Natural Sciences and Engineering Research Council of Canada,
Calcul Qu{\'e}bec, Compute Ontario, Canada Foundation for Innovation, WestGrid, and Compute Canada;
Denmark {\textendash} Villum Fonden, Carlsberg Foundation, and European Commission;
New Zealand {\textendash} Marsden Fund;
Japan {\textendash} Japan Society for Promotion of Science (JSPS)
and Institute for Global Prominent Research (IGPR) of Chiba University;
Korea {\textendash} National Research Foundation of Korea (NRF);
Switzerland {\textendash} Swiss National Science Foundation (SNSF);
United Kingdom {\textendash} Department of Physics, University of Oxford.

\appendix
\restartappendixnumbering
 
\section{Effective Area}\label{sec:appendixA}

\begin{figure*}[t]
\centering
\includegraphics[width=0.5\linewidth]{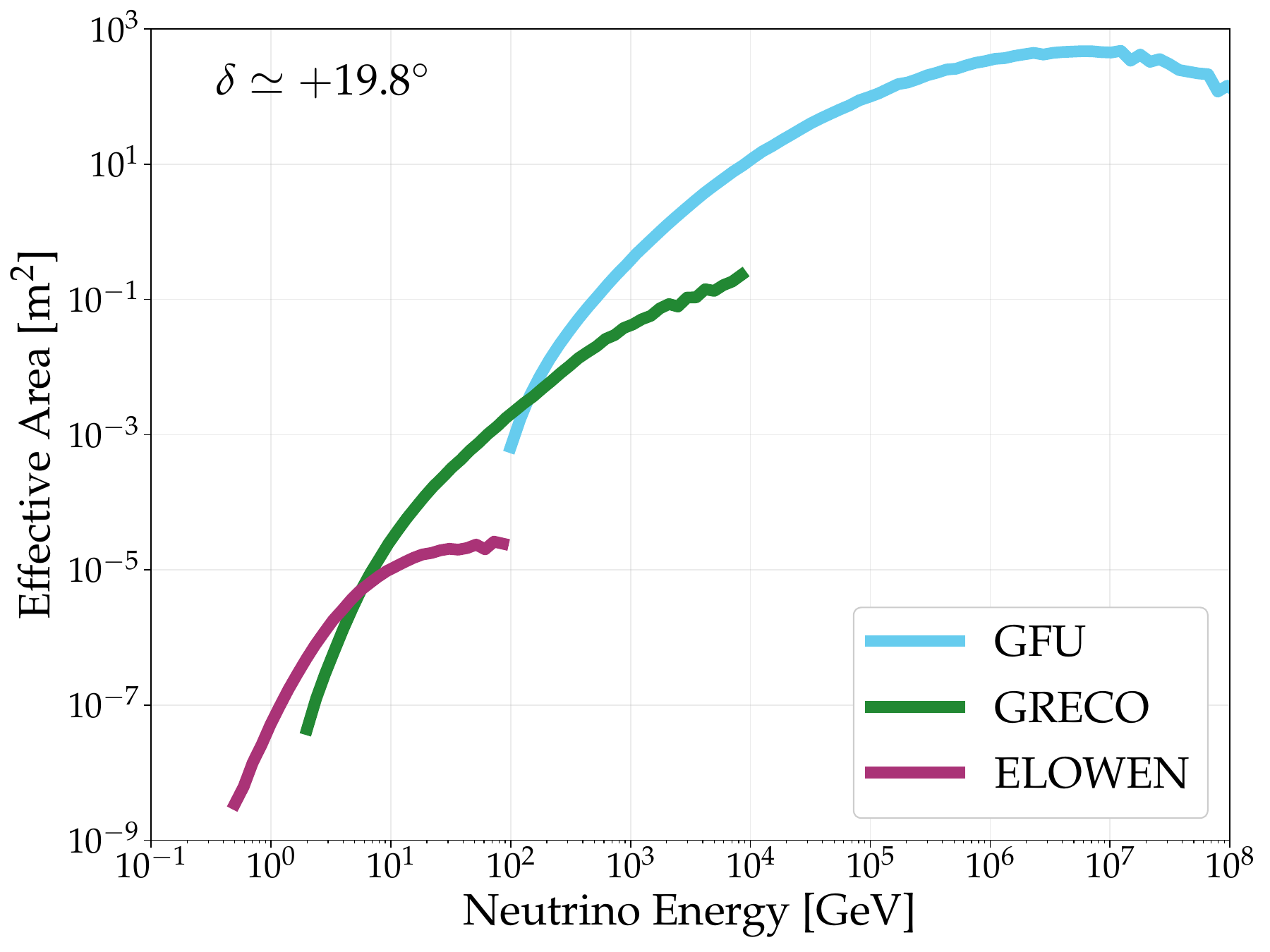}\\[-0.2cm]
\caption{The all-flavor $\nu+\bar{\nu}$ averaged effective area for neutrinos from the direction of GRB\,221009A for various datasets used in this Letter. The GRECO and ELOWEN samples are sensitive to all neutrino flavors, while the GFU sample only includes muon-neutrino-induced muons that traverse the detector as signal events.}
\label{fig:effective_area}
\end{figure*}

The complementary energy coverage of the different data sets used in this Letter can be appreciated by a comparison of the respective neutrino effective areas. The declination and energy-dependent all-flavor ${\nu+\bar{\nu}}$ averaged effective area $A_{\rm eff}(\delta, E_\nu)$ is defined via the relation:
\begin{equation}\label{eq:effarea}
N_{\rm events} = \int dE_\nu A_{\rm eff}(\delta,E_\nu) F_{\nu+\bar{\nu}}(E_\nu)\,,
\end{equation}
where $N_{\rm events}$ is the analysis-specific number of signal events and $F_{\nu+\bar{\nu}}(E_\nu)$ the per-flavor time-integrated neutrino flux (assuming identical flux for each flavor) for a source at declination $\delta$. Figure~\ref{fig:effective_area} shows these effective areas for the ELOWEN, GRECO, and GFU samples for the declination $\delta \simeq+19.8^\circ$ of GRB\,221009A. Starting at low energy, the ELOWEN sample contributes the most up to neutrino energies of 5\,GeV; the GRECO selection performs best in the 10-100\,GeV range; the GFU sample dominates beyond a few 100\,GeV. The GFU data sample used for our analysis is described in full by~\cite{IceCube:2016xci}. Note that the effective area of this GFU selection is substantially larger, especially at low energies, than the small subsample of GFU events used to trigger alerts~\citep{Blaufuss:2019fgv} and is similar to those from other neutrino source searches at high energy. The SNDAQ effective area is not shown, as it is does not work with individual neutrino events and so its effective area is not comparable to the other analyses and is not a good guide to sensitivity. The SNDAQ analysis instead searches for a sudden increase of the overall background rate in the detector induced by a burst of sub-GeV neutrinos. The figure of merit is not the total number of signal events, as in Eq.~(\ref{eq:effarea}), but rather the rate of correlated single-photon hits across the detector, which is a function of both flux and energy spectrum. More details on SNDAQ sensitivity can be found in \cite{IceCube:2011cwc}.

\end{document}